\documentstyle{article}
\begin{document}
\title{ Relativity of Space-Time Geometry}
\author{L.V.Verozub}
\maketitle
\centerline{  Department of Physics and Astronomy, Kharkov State
University,}
\centerline{ Kharkov 310077, Ukraine}
\centerline{E-mail: verozub@kharkov.ua}
\begin{abstract}
{We argue that space-time geometry is not absolute with respect to
the frame of reference being used. The space-time metric differential
form $ds$ in noninertial frames of reference (NIFR) is caused by the
properties of the used frames in accordance with the  Berkley -
Leibnitz - Mach - Poincar\'{e} ideas about relativity of space and time .
It is shown that the Sagnac effect and the existence of inertial forces
in NIFR can be considered from this point of view.   }
\end{abstract}

\section{Introduction}

The geometrical properties of space-time can be described only by
means of measuring instruments. At the same time, the description of the
properties of measuring instruments, strictly speaking, requires knowledge
of space-time geometry. One of the implications of it is that the
geometrical properties of space and time have no experimentally
verifiable significance by themselves but only within the aggregate
"geometry + measuring instruments". We got aware of it owing to Poincar\'{e}.
It is a development of the idea going back to Berkley \cite{Berkley} ,
Leibnitz \cite{Leibnitz} and  Mach \cite{Mach} . (Leibnitz, for
example, considered  that space and place are
abstractions from relations of ordinary objects and should be analyzed
in these terms).

If we proceed from the conception of relativity of space and time in
Berkley - Leibnitz - Mach - Poincar\'{e} (BLMP) sense , we should assume
that there is no way of quantitative description of physical phenomena
other than attributing them to a certain frame of reference which in
itself is a physical device for space and time measurements. But then the
relativity of the geometrical properties of space and time mentioned above
is nothing else but relativity of space-time geometry with respect to
the frame of reference being used.

Thus, it should be assumed that the concept of frame of reference as a
physical object whose properties are given and independent of the
properties of space and time is approximate, and only the aggregate
"frame of reference + space-time geometry" has a sense.

The Einstein theory of gravitation demonstrates relativity of
space-time with respect to distribution of matter. However , space-time
relativity with respect to measurement instruments hitherto has not
been realized in physical theory. In this paper an attempt to
show that there is also space-time relativity to measurement has
been undertaken. (See also \cite{Verozub1} ).

In our analysis of the problem we start from the fact that an
important distinction exists between a frame of reference (as  a
physical device) and a coordinate system (as  a way of the
space-time points parametrization). Any coordinate
transformation in pseudo - Euclidean space-time (when the curvature tensor,
certainly, remains equal to zero )  does not mean yet a transition from
an inertial frame of reference to the noninertial one.

At present we do not know how the space-time geometry in inertial frames
of reference (IFR) is connected with the frames properties. Under the
circumstances, we simply postulate (according to special relativity) that
space - time in IFR is pseudo-Euclidean. Next, we find the space-time
metric differential form in noninertial frames of reference (NIFR) from
the viewpoint of an observer in a NIFR who proceeds from the relativity
of space and time in the BLMP sense .

Then it appears that there are certain reasons to suppose that metric
properties of the space-time in the NIFR do not have a physical meaning in
themselves. The metric differential form $ds$ is completely conditioned by
the properties of the frame being used as is to be expected according to
the idea of relativity of space and time in the BLMP sense.

\section{The Metric Form $ds$ in NIFR.}

By a noninertial frame of reference  (NIFR)
we mean the frame , whose body of reference is formed by the point masses
moving in  the IFR under the effect of a given force field.

  It would be a mistake to identify "a priori" the transition  from an
IFR to  the  NIFR  with  the  transformation  of  coordinates  related   to
the frames. If we  act  in  such  a  way,  we   already  assume  that  the
properties of the space-time in both frames are  identical. However,for  an
observer in the NIFR , who proceeds from the relativity of space and time
in the BLM sense, space-time geometry is not given "a priori" and must be
ascertained from the analysis of the experimental data.

     We shall suppose that the reference body of the IFR or NIFR is formed
by the identical point masses $m$. If the observer is  at rest in one of the
frames , his world line will coincide with the world line of some
point of the reference body. It is obvious to the  observer  in  the  IFR
that the accelerations of the point masses forming the reference body  are
equal to zero. Certainly, this fact takes place  also  in  relativistic
sense. That is, if the differential metric form of space-time in  the  IFR
is denoted by $d\eta$  and $\zeta_{0}^{\alpha } = dx^{\alpha }/d\eta$
is  the 4- velocity vector of the   point
masses forming  the  reference  body, then the absolute derivative of  the
vector $\zeta_{0}^{\alpha }$  is equal to zero,i.e.
\begin{equation} \label{3/1 }
     D \zeta_{0}^{\alpha } /d \eta = 0  .
\end{equation}
(We mean that arbitrary coordinate system is used).

    Does this fact take place for an observer in NIFR ? That  is,  if  the
differential metric form of space-time in the NIFR is denoted by $ds$ , does
the  4-velocity  vector  $\zeta ^{\alpha } = dx^{\alpha } /d s $
  of  the  point  masses  forming  the
reference body of this NIFR obey  the equation
\begin{equation}    \label{3/2}
    D \zeta^{\alpha } /ds = 0  ?                            %2
\end{equation}

     The answer  depends  on  whether  space  and  time  are  absolute  in
Newtonian sense or they are relative in the BLMP  sense .

    If space and time  are  absolute  ,  the  point  masses  of  the  NIFR
reference body are at relative rest . A notion of relative acceleration
can be determined in a covariant way \cite{Dehnen}. This value is equal to
zero. However, eq.(\ref{3/2}), strictly speaking, are not satisfied.

     If space and time are relative  in the BLMP  sense, then for observers
in the  IFR and  NIFR  the motion of the point masses forming the
reference body (RB), which are kinematically equivalent, must be dynamically
equivalent too ( both in the nonrelativistic and relativistic sense). That
is , if from the viewpoint of the observer in the IFR,  the  point  masses
forming the NIFR RB are at rest ( are not subject to the
influence
of forces either ),then from the viewpoint of the observer in  the  NIFR
the point masses forming the RB of his frame are at rest too
(are not subject to the influence of forces either ). In other words,
if for the observer in the IFR the world lines of the IFR RB
points are, according to eq.(\ref{3/1 }), the geodesic lines, then for the
observer
 in the NIFR the world lines of the NIFR RB points also are
the geodesic lines in his space-time, which can be expressed by
eq.(\ref{3/2}). The diferential equations of these world lines at the
same times are the Lagrange equations of motion of the NIFR RB
points. It is obvious that this equations are equations of the geodesic
lines in
space-time whose metric differential form is given by  $ds=k\ dS$, where
$S$ is the Lagrange action describing the motion of the identical
material point masses $m$ forming NIFR RB , in the IFR, and $k$ is
the constant . This constant  $k=-(mc)^{-1}$, which can be seen from the
analysis of the case when the frame of reference is inertial.

 Thus, if we start from relativity of space and time in the BLM sense,
then the differential metric  form  of  space-time  in  the  NIFR  can  be
expected to have the following form
 \begin{equation}  \label{5/4}
                ds = -(mc)^{-1} \; dS ,                                %4
\end{equation}
where $S$ is the Lagrangian action of the identical point masses $m$ ,
 forming
the body of reference of the NIFR.

So, the properties of space-time in the NIFR are entirely determined by the
properties of the used frame in accordance with the idea of
relativity  of space and time in the BLMP sense.

     Let us consider two examples of the NIFR.

     1. The motion of the point masses forming the body  of  reference  is
described in the Cartesian coordinates by the Lagrange function
\begin{equation}  \label{5/5}
              L = - mc^{2}\  (1-v^{2} /c^{2} )^{1/2} + mwx,            %5
\end{equation}
where $v$ is the speed of the point masses  and $w$ is a constant.The points
of the given frame move under the effect of a constant force  along  the
axis $x$ . According to eq.(\ref{3/1 }), we have
 \begin{equation}   \label{5/6}
           ds = d\eta - (w x/c^{2} ) dx^{0},                           %6
\end{equation}
where $d\eta = (c^{2} dt^{2}  - dx^{2}  - dy^{2}  - dz^{2} )^{1/2} $ .

     The reference body of this NIFR can be realized by a system of non-
interacting electric charges  in a  constant, homogeneous electric field.

     2. The motion of the point masses  forming the body of reference is
described in Cartesian coordinates by the Lagrange function
\begin{equation}  \label{5/7}
 L = - mc^{2} (1-v^2 /c^2 )^{1/2} -
 (m \Omega _{0} /2)(\dot{x}y - x\dot{y}) ,
\end{equation}
where $\dot{x} =dx/dt$ , $\dot{y} = dx/dt$ , $\Omega _{0}$  is a constant.
 The points of such a frame
rotate in the plane $xy$ about the axis $z$ with the angular frequency
\begin{equation}      \label{5/7a}
        \Omega  = \Omega _{0}[1 + (\Omega _{0}r/c)^{2} ]^{-1/2} ,     %8
\end{equation}
where  $r=(x^{2} +y^{2})^{1/2}$ .The  speed $v$  tends to $c$ when $r\to 0$.
 For the given NIFR
\begin{equation}  \label{6/8}
      ds = d\eta + (\Omega _{0} /(2c)\  (ydx - xdy).                %9
\end{equation}

     The bodies of reference of this  frame can be realized  by a system
of noninteracting  electric  charges  in a constant ,homogeneous  magnetic
field .
     In the above NIFR $ds$ is of the form
\begin{equation}  \label{6/9}
             ds = F(x,dx) ,                                          %10
\end{equation}
where   $F(x,dx) = d\eta + f_{\alpha } dx^{\alpha } , f_{\alpha }$  is a
vector-function of $x$   and
\begin{displaymath}
 d\eta = [-g_{\alpha \beta }(x)dx^{\alpha } dx^{\beta }]^{1/2}
\end{displaymath}
is the differential metric
form  of  pseudo-Euclidean  space-time  of the  IFR  in the  used coordinate
system. Therefore,  generally speaking, the space-time in NIFR is
Finslerian \cite{Rund} with the sign- indefinite
differential metric form (\ref{6/9}) , where $F$ is a  homogeneous function
of the first degree in $dx$.

\section{Sagnac effect}

We shall show that metric differential form (\ref{5/4}) does not contradict
to experimental data. First consider the Sagnac effect.

     The phase shift in the interference of two coherent light beams on  a
rotating frame was observed by Sagnac \cite{Post} .  For a  relativistic
explanation of the effect it is postulated usually
 ,that   space-time  in any frames  of
refernce  is  pseudo-Euclidean  \cite{Ashtexar}, cite{Ananden}.
The motion in NIFR  is considered as the
relative one  in absolute pseudo- Euclidean space-time.

     However, for an  \underline{isolated}  observer  in  the  rotating
 frame,  who proceeds from the notion of space and time relativity in the
BLMP sense, the observed anisotropy in the time of light propagation  (which
contradicts from his viewpoint  to the experiments of Michelson-Morley
type)
is not a trivial effect. It must  have some "internal" physical explanation.

     Consider a disk  rotating with the constant angular velocity
$\Omega$ around the  $z$ axis. Let  $r$ and $\theta$ be the coordinates,
defined  by   the equations
\begin{equation} \label{7/-}
 x = rcos(\varphi ) , y = rsin(\varphi ) , \varphi  = \theta + \Omega t .
\end{equation}                                                     %11
    In the coordinate system $(r,\theta,z,t)$ the disk
points are at rest  and
the space-time metrical differential form $ds$ in the rotating frame is of
the form
\begin{equation}  \label{7/10}
 ds = d \eta + [\Omega r^2 /(2c)] d\theta + [\Omega ^2 r^2 /(2c^2 )] dx^{0} .
\end{equation}                                                        %12
where $d\eta$ is the a pseudo - Euclidean metric form :
\begin{equation}   \label{7/11}
  d\eta^{2} = [1 - (\Omega r)^2 /c^2 ](dx^{0})^2  -(dr)^2  - r^2 (d \theta)^2
            - 2(r^2\Omega /c)d\theta \,dx^0  - dz^2 .                  %13
\end{equation}
     An ideal clock  is a local periodic process measuring the length of
its own world line $\gamma$  to a certain scale. It follows from
eq.(\ref{7/10})  that
in the coordinate system being used , the time element between two events in
the same point, measured by an ideal clock on the rotating disk is given by
\begin{equation}  \label{7/13}
            dT = c^{-1} [(g_{00})^{1/2} + f_{0}]dx^{0},                 %14
\end{equation}
where $f_{0}  = \Omega ^2 r^2 /(2c^2)$ .

Consider spatial and time measurements in the NIFR.

     First we show  how to find the spatial element and light velocity
in the rotating frame provided space-time in the frame is
pseudo-Euclidean, i.e. $ds=d\eta$.(We proceed from the covariant method of
the  3 + 1  decomposition of space-time in general relativity which goes back
to Ulman, Pirani, Dehnen and other authors \cite{Dehnen}.

     In this case for an noninertial observer at rest  the direction of
time is given by the vector of 4-velocity
$\tau ^{\alpha }  = dx^{\alpha } /d \eta$ of the disk points,
which  satisfy the equation $g_{\alpha \beta } \tau ^{\alpha } \tau ^{\beta }
=1$.

     The physical 3-space is orthogonal to the vector $\tau ^{\alpha } $.
Therefore, the
arbitrary vector $\xi^{\alpha }$  in the point $x^{\alpha }$  can be
represented  as follows
\begin{equation}  \label{8/15}
     \xi^{\alpha }  = \overline{\xi^{\alpha }} + \lambda \tau ^{\alpha },
\end{equation}
where $ \overline{\xi^{\alpha }}$ are the spatial components. Using the
orthogonality condition $ \tau _{\alpha } \overline{\xi^{\alpha }}  = 0$ ,
we find that : $\lambda  = \tau _{\alpha }\xi^{\alpha } $. Therefore,
$ \overline{\xi^{\alpha }} = h^{\alpha }_{\beta } \xi^{\beta }$,    where
$h^{\alpha }_{\beta } = \delta ^{\alpha }_{\beta } -
\tau ^{\alpha} \tau_{\beta}$
is the operator of the spatial projection of a vector
field. The spatial projection of the metrical tensor  $g_{\alpha \beta }$
is  $\overline{g}_{\alpha \beta } = -h^{\gamma }_{\alpha } h^{\delta }_{\beta }
g_{\gamma \delta }  =  \tau _{\alpha }\tau _{\beta } - g_{\alpha \beta }$.

     The spatial element $dl$ is produced by the spatial projections of the
tensor $g_{\alpha \beta }$ and  vector $dx^{\alpha }$ :
\begin{equation}   \label{8/16}
   dl = (\overline{g}_{\alpha \beta } \overline{dx^{\alpha }} \;     %17
    \overline{dx^{\beta }})^{1/2} .
\end{equation}

     The time interval between the events in the points $x^{\alpha }$  and
  $x^{\alpha } + dx^{\alpha } $ is:
\begin{equation}  \label{8/17}
   dT = c^{-1} \tau _{\alpha } dx^{\alpha }.                      %18
\end{equation}
    Eqs.(\ref{8/15}) for the vector $ \xi^{\alpha } = dx^{\alpha }$   are
\begin{equation}    \label{8/18}
 dx^{\alpha } = \overline{dx^{\alpha }} + c\,dT \,\tau ^{\alpha } .  %19
\end{equation}
     In the used coordinate system $\tau ^{\alpha } = \lambda _1
\delta ^{\alpha }_0$,    where $\lambda _1$  is $ (g_{00})^{-1/2}$ which
follows
from the equality $g_{\alpha \beta }\tau ^{\alpha }\tau ^{\beta }=1)$
is $ (g_{00})^{-1/2}$  . Next,$ \tau _{\alpha } = g_{\alpha \beta }
(g_{00})^{-1/2}$ , $ \overline{dx}^{i} = dx^{i}$  and $\overline{g}_{00}=0$ .

  Therefore, $dl = (\overline{g}_{ik}dx^i dx^k )^{1/2} $
 and  $ dT  = c^{-1} (g_{00})^{1/2}\,dx^0$ .  With the accuracy up
to $v/c$  , we have :
$dl = [ (dr)^2 + r^2\,(d\theta)^2]^{1/2}$    and $dT =dt$.

     The equation $ d\eta = 0$   for light can be written as
\begin{equation}  \label{8/_}
g_{\alpha \beta } \overline{dx^{\alpha }} \; \overline{dx^{\beta }} +     %20
2 g_{\alpha \beta }\,c\,\tau _{\alpha }\, \overline{dx^{\beta }} +
c^2 g_{\alpha \beta }\tau ^{\alpha }\tau ^{\beta } = 0.
\end{equation}

 In virtue of the equation  $ \tau _{\alpha } \overline{dx}^{\alpha } = 0$
and  eqs.(16)  and (18) , this equation can be written as
$c^2\,d\tau^2 - (dl)^2 = 0$. Therefore, the
speed of light is $dl/d\tau  = c$.

      Let us return to the Finslerian space-time.
      For an observer in the rotating frame the time direction is given
by the vector of 4-velocity $\zeta _{\alpha } = dx^\alpha /ds$ of the disk
points , which
satisfies the equation $F(x,\zeta) = 1$.
     We shall  suppose that the spatial vector $\overline{dx^{\alpha }}$
is orthogonal to the vector $\zeta{\alpha }$  in the sense of the Finslerian
metric:
\begin{equation} \label{9/19}
 \zeta_{\alpha }\overline{dx^{\alpha }}  = 0 ,
\end{equation}
where \cite{Rund}
\begin{displaymath}
   \zeta_{\alpha } = F(x,\zeta)\,\partial  F(x, \zeta) / \partial \zeta .
\end{displaymath}

  Then, $\overline{dx ^{\alpha }}  = H^{\alpha }_{\beta } dx^{\beta }$  ,
where the operator of the spatial projection
\begin{equation}    \label{9/20}
   H^{\alpha }_{\beta } = \delta ^{\alpha }_{\beta }  -
   \zeta^{\alpha } \zeta_{\beta } .
\end{equation}

     The spatial projection of the tensor  $g_{\alpha \beta }$ is given by
\begin{equation}   \label{9/21}
  \overline{g}_{\alpha \beta } = - H^{\gamma }_{\alpha }              %22
   H^{\delta }_{\beta } g_{\alpha \beta }  .
\end{equation}
     The time interval between events in the points $x^{\alpha }$  and
$x^{\alpha } + dx^{\alpha }$   is of the form
\begin{equation}         \label{9/22}
    dT = c^{-1} \zeta_{\alpha }dx^{\alpha } .                          %23
\end{equation}

     The spatial element in NIFR is defined  by virtue of the metric form
(\ref{6/9}) and  the spatial projections  of the tensor  $g_{\alpha \beta }$
and
vector  $f_{\alpha } = g_{\alpha \beta } f^{\beta }$ as follows
\begin{equation}     \label{9/23}
   DL = (\overline{g}_{\alpha \beta } \overline{dx^{\alpha }}\;          %34
   \overline{dx^{\beta }} )^{1/2} +
   \overline{f}_{\alpha } \overline{dx^{\alpha }} .
\end{equation}

     The vector field  $\zeta^{\alpha }$   can be written as
\begin{equation}                 \label{9/24}
   \zeta^{\alpha } = \tau ^{\alpha } ( ds/d \eta)^{-1} =             %25
   (1 + f_{\beta } \tau ^{\beta })^{-1}\, \tau ^{\alpha } .
\end{equation}

     Therefore, with accuracy up to  $v/c$, the spatial element is given by
\begin{equation}  \label{9/25}
    DL = dl + f_{i}\,dx^{i} = dl(1+f_{i}k^{i}),
\end{equation}                                                      %26
where $k^i$   is the unit vector of the direction in the 3-space.
Then, from the equations $ g_{\alpha \beta }dx^{\alpha }dx^{\beta }  = 0$
and  $dx^{\alpha } = overline{dx^{\alpha }} + c\,dT\, \zeta^{\alpha }$  we
obtain
\begin{equation}     \label{9/26}
  g_{\alpha \beta } \overline{dx^{\alpha }} \; \overline {dx^{\beta }} +
  2 g_{\alpha \beta }\overline{dx^{a}}\,c\,\zeta^{\alpha}\,dT +
  c^2 (dT)^2 g_{\alpha \beta } \,d\zeta^{\alpha }\,d\zeta^{\beta } = 0 .
\end{equation}

  The first term in the left-hand side of  eq.(\ref{9/26}), with accuracy up
to $v/c$, coincides with $ -dl^2$  and the third term - with  $ c^2 (dT)^2$ .
The second term is not zero  since the orthogonality condition in the
Finslerian space-time has the form (\ref{9/19}), i.e.
\begin{equation} \label{10/27}
   g_{\alpha \beta } \zeta^{\alpha }\: \overline{dx^{\beta }}\:            %28
   (g_{\alpha \beta } \zeta^{\alpha } \zeta ^{\beta })^{-1/2} +
    f_{\alpha } dx^{\alpha } = 0.
\end{equation}

Therefore, with accuracy up to $v/c$, the  second term equals
$-2c\,dT\,f_{i}\,dx^{i}$  .
Putting  $v_{p}  = dL/dT$  and $v^{i}_{p}  =
\overline{dx^{i}} /dT =  k^{i} v_{p}$   ,  and  using  eq.(25)
(where  $f_{i} k^{i}$  is of the order of $v/c$ ), equation (26) ,
accurate up to $v/c$ , can be written as
\begin{equation}   \label{10/-}
    v^{2}_{p} + 2v_{p} c f_{i} k^{i} - c^2 = 0 .                       %29
\end{equation}
The solution of this equation, which coincides with the light  velocity  c
in noninertial frames, is given by
\begin{equation} \label{10/28}
    v_{p} = c (1 - f_{i} k^{i})  .
\end{equation}

 In virtue of  equations (\ref{9/25}) and (\ref{10/28})  the time of light
spread
from the point $x^{i}$  to $x^{i}  + dx^{i}$  is
$ dL/v_{p}  =  c^{-1}dl(1 + 2*f_{i} k^{i} )$  . The
unique nonzero component of the 3-vector $f_{i}$   is $\Omega r^2 /(2c)$.
(See eq.(\ref{7/10})).
For this reason the difference in time interval between light propagation
around  the rotating  disk in a clockwise and counterclockwise direction is
 $4 \pi r^2 \Omega / c^{2}$  , which
gives the  Sagnac phase shift \cite{Post}.
     Thus, the Sagnac effect for the isolated observer in the rotating
frame can be treated as caused by the Finslerian metric of space-time  in
noninertial frames of reference , which conditions the anisotropy
of  the space element and the velocity of light.

\section{ Inertial Forces}
     Let us show that the existence of the inertial forces in NIFR can be
interpreted as the exhibition of the Finslerian connection of space-time
in  such frames .
    According to our initial assumption in Section 2 ,the differential
equations of motion in an IFR of the point masses , forming the
reference body of the NIFR,are the geodesic lines of space-time in NIFR.
These
equations  can be found from the variational principle $\delta \int ds = 0$ .

The equations are of the form
\begin{equation}      \label{11/29}
     du^{\alpha }/ds + G^{\alpha }(x,u) = 0 ,                          %31
\end{equation}
where  $u^{\alpha }$ is the 4-velocity of the point mass, the world line of
which is $x^{\alpha } = x^{\alpha }(s)$, and
\begin{equation}   \label{11/-}
 G^{\alpha }(x,u) = \Gamma ^{\alpha }_{\beta \gamma } u^{\alpha }u^{\gamma }
 + B^{\alpha }_{\beta } u^{\beta } + u^{\alpha }\, \beta\, d(\beta ^{-1})/ds,
\end{equation}
where
\begin{displaymath}
  \beta = d\eta /ds = (g_{a\beta } u^{\alpha }u^{\beta })^{1/2},\;
  B^{\alpha }_{\beta } = g^{\alpha \delta } B_{\delta \beta }
\end{displaymath}  and
\begin{displaymath}
  B_{\delta \beta } = \partial f_{\beta }/ \partial x ^{\delta}  -
  \partial f_{\delta } / \partial x^{\beta } .
\end{displaymath}

In the Finslerian space-time a number of connections can be defined
according to
eq. (\ref{11/29}) \cite{Rund}. In particular, this equation can be interpreted
in
the sense that in the NIFR space-time the absolute derivative of a vector
field $\xi^{a}(x)$ along the world line $x^{\alpha } = x^{\alpha }(s)$ is
of the form
\begin{equation}   \label{11/30}
  D \xi^{\alpha }/ds  = d\xi^{\alpha } /ds +
  G^{\alpha }_{\beta } (x,dx/ds) \xi ^{\beta },                       %33
\end{equation}
where
\begin{displaymath}
 G^{\alpha }_{\beta }(x, dx/ds) =
\Gamma ^{\alpha }_{\beta \gamma } dx^{\gamma }/ds
+ \beta \, d \beta ^{-1}/ds .
\end{displaymath}

Equation (\ref{11/30}) defines a  connection of Laugvitz type
\cite{Rund} in space-time of the NIFR , which is nonlinear relative to
$dx^{\alpha }$ . The change in the vector $\xi^{\alpha }$
because of an infinitesimal parallel transport is
\begin{equation} \label{11-1}
        d\xi^{\alpha }  = - G^{\alpha }_{\beta }(x,dx) \xi^{\beta } ,
\end{equation}

    Consider  the motion of a particle of  the  mass  $m$ in a NIFR
unneffected by forces of any kind in the laboratory (inertial) frame of
reference. The differential equations of motion of  such a particle can be
found from the variational principle $\delta \int d \eta  = 0$.
Since $ds = d\eta - f_{\alpha } dx^{\alpha }$ ,
the equations  of motion are
\begin{equation}  \label{11/31}
        Du^{\alpha } /ds = B^{\alpha }_{\beta } u^{\beta } .          %34
\end{equation}

    As an  example, consider   the nonrelativistic  disk rotating in the
$xy$  plane  about the $z$ axis  with the angular velocity $\Omega$  .
Under the circumstances  the equations of motion (\ref{11/29}) are
\begin{equation}  \label{11/32}
          d \vec{v}/dt + \vec{\Omega} \times \vec{r}  = 0 ,          %35
\end{equation}
where $\vec{v} = d\vec{r}/dt$ ,  $\vec{r} = \{x,y,z\}$ and the coordinates
origin  concides with the disk center.
     The absolute derivative (\ref{11/30}) of the vector $\vec{\xi}$  is
given by
\begin{equation}   \label{12/33}
      D\vec{\xi}/dt = d \vec{\xi} /dt - \vec{\Omega } \times \vec{ \xi}.
\end{equation}
and the equations of motion (\ref{11/31}) of the considered particle in the
NIFR are
\begin{equation}     \label{12/34}
      D \vec{v} /dt = - \vec{\Omega } \times \vec{ v} .
\end{equation}
Next, for the 4-velocity  $u^{\alpha }$  we have                       %38
\begin{equation}   \label{12/35}
      u^{\alpha }  = \overline{u}^{\alpha }  + \lambda \zeta^{\alpha } ,
\end{equation}
where $\lambda  = \zeta_{\alpha } u^{\alpha }$  , $\overline{u}^{\alpha }$
  is the velocity of the particle in the NIFR found
with the help of measuring  instruments. In the nonrelativistic limit
eq.(\ref{12/35}) is written in the form
\begin{equation}  \label{12/36}
      \vec{v} = \overline{\vec{v}} + \vec{\tau } ,
\end{equation}
where $\overline{\vec{v}}$ is the relative velocity of the particle and
$\vec{\tau }$ is the velocity of
the disk point in the laboratory frame.
     Substituting (\ref{12/36}) in (\ref{12/33}), we find that
\begin{equation}   \label{12/37}
        D \overline{\vec{v}}/dt = - D\overline{\vec{\tau }} /dt         %40
 - \vec{\Omega } \times \vec{v} - \vec{\Omega } \times \vec{\tau }.
\end{equation}

     The value $ D\overline{\vec{v}}/dt$  is an acceleration of the
considered particle in
the used NIFR found with the help of measuring instruments.
    The velocities  field $\vec{\tau}$  of the disk point is given by
$\vec{\tau } =
 \vec{\Omega } \times \vec{r}$.
Hence, along the particle path we have
$d\vec{\tau } /dt = \vec{\Omega} \times \vec{v}$ and                    %41
\begin{equation}     \label{12/38}
  D \vec{\tau } /dt = d\vec{\tau }/dt - \vec{\Omega} \times \vec{\tau } =
  \vec{\Omega } \times \overline{\vec{v}} .
\end{equation}
    Thus, finally, we find from (\ref{12/34})
\begin{equation}   \label{12/39}
     mD\overline{\vec{v}}/dt = -2m(\vec{\Omega } \times \overline{\vec{v}}) -
 m \vec{\Omega } \times ( \vec{\Omega} \times \vec{r}) .
\end{equation}                                                         %42
    We  arrived at the nonrelativistic equations of motion of a point in a
rotating frame \cite{Syng}. The right-hand of eq.(\ref {12/33}) is the
ordinary expression for the Coriolis forces and the centrifugal force in the
rotating frames. (See also
\cite{Verozub} and   \cite{Tavakol} ).

     Thus, in the nonrelativistic limit the Finslerian space-time in NIFR
manifests itself in the structure of vector derivatives with respect
to time $t$ .
     It should be noted that eq.(36) is considered sometimes
in classical dynamics nominally \cite{Syng} just for the derivation
of the inertial forces in NIFR's.

\section{Experimental test}
Consider an  experimentally verifiable consequence of the
above theory.

Let $p^{\alpha} = mc\ dx^{\alpha} / d \eta$ be 4-momentum of a
particle in the IFR. Using $3+1$ decomposition of space-time in
the NIFR we have
\begin{equation}
p^{\alpha} = \overline{p}^{\alpha} + E \zeta^{\alpha}
\end{equation}
where $\zeta^{\alpha} = dx^{\alpha} /ds $ . From the viewpoint
of an observer in the NIFR the spacial projection
$ \overline{p}^{\alpha}$ should be identified with the momentum
, and the quantity $cE$ with the energy ${\cal E}$ of the
particle.  It is obvious that $E=\zeta_{\alpha} p^{\alpha}$,
where $\zeta_{\alpha}$ is defined in eq. (20) from Sec.2 .

Therefore, the energy of the particle in the NIFR is
\begin{equation}
{\cal E} = m Q c^2 \zeta_{\alpha} u^{\alpha} ,
\end{equation}
where $Q=ds/d \eta = F(x, dx/d \eta)$ and $u^{\alpha} =
dx^{\alpha} / ds$ is the 4-velocity of the particle. For the
particle at rest in the NIFR $u^{\alpha}= \zeta^{\alpha}$
and we obtain
\begin{equation}
{\cal E} = mQc^2
\end{equation}
Thus, the inertial mass $m_{n}$ of the particle in the NIFR is
given by
\begin{equation}
m_{n} =Qm
\end{equation}
The quantity $m_{n}$ coincides with the proportionality factor
between the momentum $\overline{p}^{\alpha}$ and the velocity
$\overline{v}^{\alpha} = \overline{dx}^{\alpha}/dT$ of a
nonrelativistic particle in the NIFR.

Since $Q$ is the function of $x^{\alpha}$ , the inertial mass in
the NIFR is not a constant. For example, on the rotating disk
we have
\begin{equation}
m_{n} = m\  /(1-\Omega^{2}r^{2}/2c^{2}),
\end{equation}
where $\Omega$ is the rotation angular velocity  and $r$ is the
distance of the body from the disk center.

The difference between the inertial mass $m_{e}$ of a body on the Earth's
equator and the mass $m_{p}$ of the same body on the pole is given
by
\begin{equation}
(m_{e} - m_{p})/m_{p} = 1.2 \cdot 10^{-12}
\end{equation}
The dependence of the inertial mass of particles on the Earth's longitude
can be observed by the M$\ddot{o}$ssbauer effect. Indeed, the change
$\Delta \lambda $ in the wave length $\lambda $ at the Compton
scattering on particles of the masses $m$ is proportional to $m^{-1}$.
If this value is measured for  $\gamma $ quantums with the help of
the M$\ddot{o}$ssbauer effect at a fixed scattering angle, then after
transporting the measuring device from the longitude $\varphi $ to
the longitude $\varphi_{1}$ we have
\begin{equation}
\frac{ (\Delta \lambda )_{\varphi }^{-1} - ( \Delta \lambda )_{\varphi_{1} }
^{-1} } { (\Delta \lambda )_{\varphi }^{-1} } =
K\ [ cos(\varphi )^2 - cos(\varphi_{1} )^2 ],
\end{equation}
where $K$ is a constant.

\section{ Conclusion}.

     The above theory gives in some way the realization of the idea of
space-time properties relativity in the BLMP  sense.
Starting from the given properties of space-time in the IFR , the theory
demonstrates that space-time geometry in the NIFR is  caused by the
properties of
the employed frames of reference.

     The existence of inertial forces in NIFR obviously indicates that
Newtonian inertia principle is violated in such  frames. For this reason
the Einstein's general relativity principle (we mean his assumption  that
physical laws are identical in any frames of reference ) is based on
the interpretation of inertial forces as the "exterior" ones (in Mach sense).
Our analysis of  the  problem  shows the deep connection between the
existence of inertial forces in NIFR and space-time  geometry in such frames.

\end{document}